\documentstyle[12pt]{article}
\def\a{\begin{eqnarray}}
\def\b{\end{eqnarray}}
\def\0{\nonumber}
\def\ba{\begin{array}}
\def\ea{\end{array}}

\def\cL{{\cal L}}
\def\sD{\slash\hskip-9pt D}

\begin{document}
\setcounter{page}{1}
\begin{center}
\hfill    SISSA 161/97/EP/FM\\
\hfill    hep-th/9712205\\
\vskip .2in
{\LARGE {\bf Anomalies and Locality in Field Theories and M--theory  }}
\vskip .2in
{ \large L. Bonora$^{a,b}$, C.S. Chu $^{a}$, M.Rinaldi $^{c}$}
{}~\\
\quad \\
{\em ~$~^{(a)}$ International School for Advanced Studies (SISSA/ISAS),}\\
{\em Via Beirut 2, 34014 Trieste, Italy}\\
{\em ~$~^{(b)}$  INFN, Sezione di Trieste}\\
{\em ~$~^{(c)}$  Dipartimento di Matematica, Universit\'a di Trieste\\
P.le Europa 1, 34127 Trieste, Italy}\\
\vskip .1in
E-mail: bonora@sissa.it, cschu@sissa.it, rinaldi@uts.univ.trieste.it
\end{center}
\begin{abstract}
{We review some basic notions on anomalies in field theories and
superstring theories, with particular emphasis on the concept of locality. 
The aim is to prepare the ground for a discussion on anomalies in theories 
with branes. In this light we review the problem of chiral anomaly cancellation
in M--theory with a 5--brane.}
\end{abstract}

\section{Introduction}

Local chiral anomalies represent a breaking of classical
infinitesimal gauge symmetries. They are an obstruction to defining path 
integrals involving chiral fermions or (anti)self--dual tensor fields.
The corresponding theories are ill--defined. Therefore one of the important 
issues in constructing a theory is to verify that it is free of these
anomalies. Once the absence of chiral perturbative anomalies is ascertained, 
there may still be global anomalies, i.e. breaking of the symmetry with respect
to gauge transformations not connected to the identity. 
One has to make sure that the theory is free
of these anomalies too. In this contribution we will be concerned with
chiral local perturbative anomalies. Historically, these anomalies have been
discovered in Yang--Mills theories, subsequently extended to gravity theories
and finally studied in (super)string theories. In the latter context they
became a fundamental tool in selecting, out of infinite many,
five superstring theories in 10D (a type I, two type II and two heterotic).

After the second string revolution, anomalies are anew an important tool in 
discriminating among different theories. The new input, with respect to
field theories or superstring theories, is the presence of branes, i.e. 
topological defects over which new degrees of freedom are defined, 
which have their own autonomous dynamics, but are also in interaction 
with the ambient theory. In this new situation we may have three new 
contributions to chiral anomalies, if the brane field content is chiral:
the anomalies of the brane in isolation, the anomalies induced via the embedding
of the brane in the ambient theory, and the inflow anomalies, which come from
the coupling of the brane to the ambient theory. The cancellation of the
anomaly resulting from the sum of all the contributions is then a basic 
criterion for consistency. 

At first sight it may look like we are improperly mixing things. After all we 
are talking about local perturbative anomalies in a non--local and 
non--perturbative setting such as the embedding of a topological defect in 
a given theory. However, first of all the term `perturbative', referred to the
chiral anomalies we are dealing with, is correct in the sense that they
can be evaluated in perturbative field theory, but it may be misleading. 
Actually these anomalies represent an incurable disease of the theory, which
is not limited to its perturbative consistency. Second, the branes we are
considering preserve part of the supersymmetry, therefore they are stable 
as long as supersymmetry survives; this means in particular that their massless
spectrum (on which the anomalies are calculated) is also stable, consequently
if the corresponding theory is anomalous it is hopelessly inconsistent: there
is no possibility that we have miscalculated the anomaly by considering 
an unstable massless spectrum. 

In this paper we discuss one of these problems, perhaps the most 
relevant one, that is the anomaly of the 5--brane embedded in M--theory.
Our aim here is to view this problem in the framework of the principle 
of {\it locality} in field theory. To this end we start from very basic
notions of anomalies in field theory. Section 2 contains a pedagogical review
of perturbative chiral anomalies in field and string theories with 
particular emphasis on locality. In section 3 we collect the information
we need on M--theory and the M--theory 5--brane in order to discuss the
relevant anomaly problem. The M--5--brane `naked eye' anomaly does not 
identically vanish, so one has to set out to find some sophisticated
cancellation mechanism. In section 4 we present two such cancellation 
mechanisms, one is of topological nature while the other is provided by a 
local counterterm. We discuss the problems that arise if we accept
these mechanisms.

\section{Chiral perturbative anomalies in field theories}

Let us consider a gauge theory with gauge group $G$ in which chiral fermions 
$\psi$ are
coupled to the gauge potential $A$. The fermions belong to a given 
representation of the Lie algebra ${\cal G}$ with generators $T^a$,
$a=1,...,{\rm dim} G$, and $A= \sum_a A^aT^a$. The classical action is of
the form 
\a
S  = \int \bar \psi~ {\sD}_\chi(A)\psi, \0
\b
where ${\sD}_\chi(A)$ is the chiral (left or right) covariant Dirac operator.
$S$ is invariant under the gauge transformations
\a
\delta A = d\xi + [A,\xi],\qquad \delta \psi = \xi \psi, \0
\b
where $\xi =\sum_a\xi^a T^a$. Therefore we have the classical Ward--Takahashi 
identity $\delta S=0$. At one--loop $S$ is replaced by the one--particle 
irreducible vertex operator $\Gamma_1$.     
$\Gamma_1$ may not satisfy the Ward--Takahashi identity. In general one finds
\a
 \delta \Gamma_1 = \hbar {\cal A} + O (\hbar ^2). \0
\b
${\cal A}$ turns out to be a local functional of the fields, i.e. an
integrated polynomial 
of the fields and their derivatives, linear in $\xi$. The functional 
operator $\delta$ can be made into a coboundary operator: $\delta^2 =0$.
Then ${\cal A}$ is seen to satisfy the {\it Wess--Zumino consistency
conditions} \cite{WZ}
\a
\delta {\cal A}=0. \0
\b
Now one is faced with two possibilities. Either 
\a
{\cal A} =\delta {\cal C} \0
\b 
for some local functional of the fields ${\cal C}$, or this is not possible
for any ${\cal C}$. In the first case the classical Ward--Takahashi identity
can be reconstructed at one--loop by suitably redefining $\Gamma_1$, as follows:
\a
\hat \Gamma_1= \Gamma_1 - \hbar {\cal C}, \qquad \delta \hat \Gamma_1 = 
O(\hbar^2). \0
\b
In the second case the classical gauge invariance is broken and we say we have 
an {\it anomaly}. 

What the presence of an anomaly means is  not only that we cannot retrieve a
classical conservation law (this in itself may not be a problem), but
that there is an obstruction to defining the determinant of the chiral
Dirac operator ${\sD}_\chi(A)$, i.e. we cannot properly define the functional
integral of the corresponding theory. 

It is therefore crucial to know whether a given theory is anomaly free or not.
This takes two steps: the characterization of the anomalies and the method
to calculate them.

\subsection{Anomalies and locality}

From the very definition we see that an anomaly corresponds
to a non--trivial cohomology class of $\delta$, i.e. an anomaly is only defined
up to $\delta {\cal C}$, for any local functional ${\cal C}$. Let us make 
a few remarks on the issue of {\it locality}. The idea of locality in field
theory is translated into the idea that the only allowed terms are 
integrals over the space--time manifold of polynomials of $A$
and of the gauge transformations $\xi$, as well as of their derivatives 
\cite{BRS}.
This definition based on polynomiality works well for Yang--Mills gauge 
theories, 
but it is not quite fit for gravity theories, where metric and vielbein fields 
are involved. We will shortly give a definition of locality that bypasses this
difficulty. But, for the time being, let us concentrate on gauge theories 
without gravity. If the space--time manifold is $M$ we can write
\a
{\cal A}= \int_M {\omega}_n^1, \quad\quad n={\rm dim}~M, \0
\b
where ${\omega}_n^1$ is an $n$--form constructed as a polynomial in $A$ and 
$\xi$
and their derivatives, but linear in $\xi$. Then we can translate 
the above conditions on ${\cal A}$ as follows:
\a
 \delta {\cal A}= 0\quad \leftrightarrow \quad
 \delta {\omega}_n^1 +d {\omega}_{n-1}^2 =0 
\label{cocy}
 \b
 for some ${\omega}_n^1,{\omega}_{n-1}^2$, and
 \a
 {\cal A}\neq \delta {\cal C}\qquad \leftrightarrow\qquad 
{\omega}_n^1 \neq \delta {\omega}_n^0 + d{\omega}_{n-1}^1\label{anom}
\b
for any ${\omega}_{n-1}^1, {\omega}_n^0$. The subscript of ${\omega}$ denotes 
the form 
degree, while the superscript represents the  number of $\xi$'s that appear in 
the expression.

The problem represented by (\ref{cocy},\ref{anom}) was solved long ago 
\cite{bc,dtv}.
The solutions, i.e. the anomalies, are given in terms of (reducible or 
irreducible) symmetric polynomials $P_k$, with $k$ entries, 
which are invariant under the adjoint action of $G$ in ${\cal G}$. Let $P_k$
be the polynomial that corresponds to a given anomaly, and let
\a
F_t = tdA + {1\over 2} t^2[A,A],\qquad 0\leq t\leq 1, 
\qquad F_1\equiv F.
\label{curv}
\b
Then we write the descent equations for $P_k$. The first is the Chern--Weil 
formula
\a
P_k(F,...,F) = d\Big( k\int_0^1dt~P_k(A,F_t,...,F_t)\Big) 
\equiv d~\Big(TP_k(A)\Big) .
\label{CW}
\b
The second determines the anomaly ${\omega}_n^1$
\a
\delta ~TP_k(A) + d {\omega}^1_n=0,\quad\quad n=2k-2 .
\label{desc}
\b
The explicit expression of the anomaly can be given as
\a
{\omega}^1_n = k(k-1) \int_0^1 dt~P_k(d\xi, A, F_t,...,F_t), 
\quad n=2k-2 .
\label{anomaly}
\b

\subsection{Locality and universality}

In the following we have to be a bit more precise. 
So we introduce the principal
fiber bundle $P(M,G)$ where our gauge theory is defined. Then $A$ is a 
connection in $P$, $F$ its curvature, $P_k(F,...,F)$ are basic forms while
$TP_k(A)$ are not, and so on.
Let us consider, for example, a reducible polynomial $P_k = P_{k_1}P_{k_2}$,
$k=k_1+k_2$, and 
suppose that $P_{k_2}(F,...,F)$ is a trivial class in $H^{2k}(M)$, i.e.
\a
P_{k_2}(F,...,F)= d \alpha, \quad \mbox{where $\alpha$ is a basic form 
in $P(M,G)$}. 
\b
This happens if $P_{k_2}(F,...,F)$ is in the kernel of the
Weil homomorphism. 
Then  the Chern-Simons form corresponding to $P_k$ can be written
as $-d (TP_{k_1}(A)\alpha)$ since $P_{k_1}(F,...,F)\alpha$ is a basic 
$n+1(=2k-1)$ form
and so identically vanishes. This Chern--Simons term will give rise,
via the descent equations, to an anomaly which is a non--trivial 
local anomaly (unless the field content of the theory is enlarged, see below)
although it is trivial from a cohomological point of view.
It was customary some time ago to call such an anomaly {\it non--topological} 
to make a distinction with the {\it topological} ones, which correspond
either to irreducible polynomials or to reducible ones in which any factor
is cohomologically non--trivial.

From the above example we see that the cohomology of $P(M,G)$ is not sufficient 
to tell us which are the local anomalies. For that we need a different kind of
information, called the {\it locality} or {\it universality principle}, 
\cite{BCRS1}.

Any principal fiber bundle, such as $P(M,G)$, can be obtained by pulling back
the universal bundle $EG$ by a suitable classifying map $f$, according
to the diagram
\a
\begin{array}{ccc}
{ P}& \buildrel {\hat f} \over {\longrightarrow} &{ EG}\\
  {\downarrow}&{}& {\downarrow}\\
{ M}&\buildrel f \over {\longrightarrow}&{ BG}
\end{array}
\b
$BG$ is the classifying space, $\hat f$ is any map that projects to $f$. In 
$EG$ there exists a universal connection $A_u$, such that any connection $A$ in 
$P$ can be obtained as $A= \hat f^* A_u$, via some $\hat f$. Now, take any 
polynomial $P_k$ and construct $TP_k(A_u)$ in $EG$. Forms in $EG$, constructed
like this out of $A_u$, will be called {\it universal}. Now we are ready
to state the {\it universality principle for anomalies}: local anomalies
are $n$--forms (modulo exact universal $n$--forms) 
characterized by (obtained from) universal $n+1$--forms $TP_k(A_u)$. 
Here $P_k$ can be a reducible or irreducible polynomial.
This principle identifies local perturbative anomalies in Yang--Mills
gauge theories and extends to gravity theories as well, \cite{BCRS1}.
It gives a rigorous formulation to the idea that in field theory local 
perturbative anomalies are {\it local}, i.e. they do not depend on the
global details of the space--time $M$.

What remains for us to recall is the method to compute anomalies once the
massless fermionic spectrum of a given theory is specified. By far the simplest
way is to use Atiyah--Singer's family index theorem. This theorem provides,
in a space $M$ of dimension $n$, the $n+2$--form from which the anomaly can be 
extracted via the descent equations, as above.

One may wonder why we worry about the locality principle when the index theorem
gives us a precise expression for all the anomalies. The point is that in most
theories we have different types of zero modes (chiral spin 1/2 and 
spin 3/2 fermions, selfdual or antiselfdual tensors, see below), each 
one characterized
by its own index theorem. Moreover, in more complex theories, there may be
inflow anomalies (see below). Of course what is important is the resulting 
anomaly, namely the sum of the anomalies corresponding to each zero mode. 
The result 
of this calculation has then to be compared with the locality principle: is the
total anomaly a true local anomaly or can it be cancelled by a local
counterterm?

\subsection{Getting around universality}

Up to now we have discussed the correct way to deal with anomalies 
in Yang--Mills and gravity theories (for a review, see \cite{ZS,AGW}). 
String theories carry something new
in this panorama: the Green--Schwarz mechanism. The low energy
effective field theory that represents the interactions of the massless modes 
of the type I superstring in 10D (let us concentrate on this theory for
simplicity) is N=1 super--Yang--Mills coupled to supergravity
theory. It is a chiral theory and it is anomalous for a generic gauge
group $G$. However, if $G=SO(32)$ or $E_8 \times E_8$,
the expression of the anomaly gets drastically simplified, all the irreducible
polynomials cancel and we are left with  the following expression
for the anomaly
\a
{\cal A} = \int_{M_{10}} \Big( {\omega}_{2L}^1- {\omega}_{2G}^1 \Big)
X_8.
\label{tI}
\b
In order to explain this equation we introduce, beside the gauge connection
$A$ with curvature $F$, also the spin connection $\omega$ with curvature $R$.
Moreover let us call $K$ the relevant polynomials with two entries -- 
essentially the (suitably normalized) Killing forms in the corresponding groups.
Then ${\omega}_{2L}^1$ is the 2--form anomaly that comes, via descent equations,
from the 4--form polynomial $K(R,R)$. Similarly ${\omega}_{2G}^1$ comes
from $K(F,F)$. The relative Chern--Simons terms are 
\a
\omega_{3L} := TK(\omega)= 2\int_0^1 dt~K(\omega, R_t),
\qquad{\rm and}\qquad
\omega_{3G} := TK(A)=2 \int_0^1 dt~K(A, F_t),\0
\b
respectively. Finally, in eq.(\ref{tI}), $X_8$ is an 8--form constructed with
$R$ and $F$, invariant under both gauge and Lorentz transformations.
The suggestion of Green and Schwarz, \cite{GS}, was that
the  2--form field $B$ in the theory and its modified curvature $H$ incorporate 
the cancellation mechanism for the residual anomaly (\ref{tI}). More precisely,
\a
d H= K(F,F)-K(R,R),\qquad H=d B+ \omega_{3G}-\omega_{3L}.
\label{GS}
\b
Therefore, if we assume $\delta H=0$, we get $\delta B = 
{\omega}_{2G}^1-{\omega}_{2L}^1$. 
Consequently the conterterm $\int_{M_{10}} B X_8$ cancels the residual 
anomaly.
 
We have reported this well--known construction here in order to relate it
to the locality principle. What we have done here is to suppose first that
\a
K(F,F)-K(R,R)= d\gamma, \quad \mbox{ where $\gamma$ is a basic form}, 
\label{gamma}
\b
i.e. that the LHS is cohomologically trivial as a form in $M$. This is exactly
the case considered at the beginning of section 2.2. We saw there that this 
by itself is not enough to trivialize the local anomaly (\ref{tI}).
In fact what is understood in (\ref{GS}) is much more than that: it means that
we have promoted $\gamma$ to a local field and identified it with $H$;
at this point it is consistent to require that $B$ is a local field
with the properties described above\footnote{Actually 
(\ref{GS}) is not accurately defined since $\omega_{3G}$ and $\omega_{3L}$
are forms in the total space of the respective principal fiber bundles, and are 
not basic, contrary to what $B$ and $H$ are supposed to be. 
However this can be taken care of by introducing suitable background 
connections \cite{BCRS1}. See also the remark at the end of section 3.},
although its geometrical nature remains to be defined.

At first sight the properties of the local fields $B$ and $H$, although
not inconsistent, 
may sound arbitrary. But the field theory we are considering here is 
the low energy effective field theory (LEEFT) of a superstring theory.
One can therefore compute the anomaly in another way: 
just using superstring theory instead of its LEEFT. 
If one does the relevant computation,
\cite{GSW}, one finds that the residual anomaly (\ref{tI}) actually does not
appear. In other words (\ref{tI}) is an artifact of the LEEFT. Therefore if
the latter is to represent the low energy dynamics of
superstring theory, it is not only allowed but
also compulsory to promote $\gamma$ to a local field and embody it in $H$, etc.
We emphasize again that the anomaly (\ref{tI}) is a genuine non--trivial
local anomaly that passes the universality test. However the `exotic'
properties of $H$ and $B$ `trivialize' it. 

The Green--Schwarz mechanism is paradigmatic. It suggests how in some
theories we can get around the universality principle. Similar and generalized  
Green-Schwarz mechanisms have been applied to many other different
circumstances. Below we are going to discuss a difficult case,
the case of anomalies on 5--branes embedded in M--theory, in which
the idea of locality is rather put under strain\footnote{Other cases of theories
involving branes seem to offer unambiguous if not straightforward cancellation 
mechanisms for their anomalies, \cite{m,yin}}. We will try, also in that
context, to stick as much as possible to what has been said in this section.

\section{The M--5--brane and its anomaly}

The M--theory anomalies can be studied by looking at the massless fields
of the LEEFT, i.e. 11D supergravity. These are the graviton, the gravitino
and the components of the 3--form $C_3$. Of course there are no perturbative 
anomalies in odd dimensions and so M--theory in isolation is perturbative
anomaly--free. 
 
The world--volume massless content of the M--theory 5--brane fills up the 
tensor multiplet of $(2,0)$ supersymmetry in 6D. It contains 8 chiral 
spinors $\psi$, 5 bosons $X^i$, $i=1,...,5$, which represent the directions 
transverse to the 5--brane and 3 components of an antiselfdual two--form 
$B_{ij}^-$. So the corresponding theory is 
potentially anomalous. However, since there are no inborn gauge field or 
metric on the 5--brane, the latter is anomaly free when in isolation.
World volume dynamics of the M--5--brane has been determined in
\cite{schwarz,pst,hs}. 

However the situation changes if the 5--brane is embedded in M--theory.
Geometrically
this corresponds to having a six--dimensional manifold $W$, the world--volume 
of the 5--brane, embedded in the eleven--dimensional manifold $Q$ of M--theory.
The anomalies in question arise from a breakdown of invariance under the 
diffeomorphisms of $Q$ that preserve the embedding of $W$ in $Q$. More 
precisely,
the tangent bundle $TQ$ of $Q$ restricted to $W$ decomposes according to
$\left. TQ\right\vert_W = TW \oplus N$, where $TW$ and $N$ are the tangent 
bundle and the
normal bundle of $W$, respectively. A Riemannian metric on $Q$ induces a 
Riemannian metric on $W$ and a metric and an $SO(5)$ connection on $N$. Now the 
relevant diffeomorphisms of $Q$ are those that map $W$ to $W$. Any such 
diffeomorphism generates a diffeomorphism of $W$; if the diffeomorphism
induced on $W$ is the identity, it determines a gauge transformation in
$N$. Therefore the anomalies to be considered are the usual gravity anomalies
in $TW$ and the gauge anomalies in $N$. They can be calculated by means of the
index theorem. Their evaluation has been carried out in \cite{witten1}.
The 8--form corresponding to the total anomaly from $\psi$ and 
$B^-$ is given by
\a
I_{B^-}+ I_{\psi}&\,=\,&-{1\over 8}\left.  L(W)\right\vert_8 + 
{1\over 2} \hat A (TW) 
\left. ch S(N)\right\vert_8
\label{anom1}\\
&\,=\,& {1\over {192} }\Big(p_1(TW)-p_1(N)\Big)^2 + {1\over{48}}
\Big(p_2(N) - p_2(TW)\Big), \0
\b
where $L$ is the $L$ polynomial, $\hat A$ is the arithmetic genus and $ch$ is
the Chern character of the spin bundle $S(N)$ of which the $\psi$ are sections, 
i.e.
the tensor product of the spin bundle over $W$ tensored with the vector
bundle associated to the normal bundle $N$ via the representation ${\bf 4}$
of SO(5). The symbol $|_8$ means that we extract the 8--form part of the
corresponding expression. $p_1$ and $p_2$ denote the appropriate Pontrjagin 
classes. This is not
the end of the story, as far as perturbative anomalies are concerned. In
the 11D supergravity action one must include also 
the gravitational correction term \cite{w1,d1}  
\a
S_{bulk} = \int_Q C_3\wedge I_8, \quad\quad
I_8 = \frac{1}{48} \Big( -{1\over 4} p_1(TQ)^2 + p_2(TQ) \Big) . 
\label{bulk}
\b
Notice that this term (\ref{bulk}) is not quite written in 
the proper way when a 5--brane is embedded. 
For supergravity in isolation we have
\a
F_4 = dC_3 ,
\b 
and so
\a 
dF_4 =0. \0
\b
When the 5--brane is present this is replaced by \cite{witten2,witten1}
\a
dF_4= \delta_W .
\label{mag}
\b
Physically this means that the 5--brane is magnetically 
coupled to M--theory. From the mathematical point of view, 
$\delta_W$ is a 5--form which can be 
interpreted as the Poincar\`e dual of the submanifold $W$ of $Q$. 
It can be chosen to have (a delta--function--like) support on $W$. 
One can now write (\ref{bulk}) in the appropriate form in the presence of
M--5--brane
\a
S_{bulk} = \int_Q F_4\wedge TI_8,
\label{bulk1}
\b
Taking its variation with respect
to the above mentioned gauge transformations and using (\ref{mag}),
we find 
\a
\delta S_{bulk} =  \delta \int_Q F_4\wedge TI_8= 
-\int_Q F_4 \wedge d{\omega}_{in} =
\int_W {\omega}_{in},
\label{A61}
\b
where ${\omega}_{in}$ is the {\it inflow} anomaly 
generated via the descent equations from the 8--form $I_8$, which can be
rewritten as
\a
I_8 = -{1\over {192} }\Big(p_1(TW)-p_1(N)\Big)^2 + {1\over{48}}
\Big(p_2(N) + p_2(TW)\Big)
\label{Ibulk}
\b
Adding up (\ref{anom1}) and (\ref{Ibulk}), we obtain the total
perturbative anomaly of the 5--brane. It is generated by the 8--form 
\a
{1\over 24} p_2(N).
\b
Let us introduce, for later use, the corresponding
properly normalized Chern--Simons term and 6--form anomaly $\omega_7(N)$ 
and ${\cal A}_6$, respectively:
\a
&&{1\over 24} p_2(N)=  d\omega_7(N),\0\\
&&0=\delta \omega_7(N) +d {\cal A}_6\label{p2N},
\b 
where $\delta$ represents the gauge transformations induced on the normal 
bundle.

It is shown in \cite{witten1} that this residual anomaly can be canceled 
by means of a suitable 6D counterterm in the case in which $Q$ is the 
product of a ten dimensional manifold and a circle, i.e. when M--theory reduces 
to type IIA superstring theory (in the limit of small radius) and the 5--brane 
reduces to the solitonic 5--brane of type IIA. In the following we want to 
discuss the anomaly cancellation problem in the general M--theory set up,
\cite{BCR}.

Before we start our analysis, let us recall another fact about anomalies.
An anomaly, being the result of a local calculation, must be represented 
by a global basic form, therefore in 
eq.(\ref{p2N}), $\omega_7(N)$ must also be a basic form.  
A global basic form $\omega_7(N)$ in $W$ can be written by introducing 
a reference connection. Let $A$ be the generic connection (with curvature $F$)
in the principal bundle $P(W, SO(5))$ to which the normal bundle
is associated, and $A_0$ the reference connection. Then 
\a
 \omega_7(N)&=& 
4\int_0^1dt~ {\rm P}_4 (A-A_0, {\cal F}_t,{\cal F}_t,{\cal F}_t)
\0\\
&=&T{\rm P}_4(A)-T{\rm P}_4(A_0) + d S(A,A_0),\label{CS1}
\b
where ${\rm P}_4$ is the fourth order symmetric $SO(5)$--ad--invariant
polynomial corresponding to $p_2(N)/24$, 
\a
{\rm P_4} (F^4) = \frac{1}{24}p_2(N).
\label{P4}
\b
$T{\rm P}_4$ is the usual Chern--Simons form (\ref{CW}) and
$S(A,A_0)$ is a suitable 6--form in $P$ (see \cite{chern}).
Here ${\cal F}_t$ is the curvature of $tA+(1-t)A_0$
and $F_t= tdA +t^2/2[A,A]$.

The three forms
in the RHS of (\ref{CS1}) are all defined in the total space of $P(W, SO(5))$,
while the LHS is basic. In general there is very little need to stress that
anomalies are basic forms, and up to now we have not done it ourselves. 
However this fact, which is usually understood, is crucial in the present 
case and must be brought to the foreground. 

For instance we remark that $\omega_7$ being basic means that it 
vanishes identically in $W$, as $W$ is a 6--dimensional space. This fact is 
immaterial as long as we deal with descent equations (for example, in
section 2 of \cite{BCRS1} it is shown how to reconcile it with the descent
equations), but becomes very important if 
one wants to use $\omega_7(N)$ to construct an action term. This will be one 
of our main concerns in the following.

\section{Mechanisms for the anomaly cancellation}

Let us start by recalling that a 5--brane magnetically couples to M--theory
via eq.(\ref{mag}). As long as (\ref{mag}) is a defining 
equation for the 5--brane, it implies, by Poincar\`e duality, that 
$W$ is the boundary of some seven--dimensional manifold $Y$ in $Q$,
\a
W=\partial Y. \label{Y}
\b

In the following we are going to exploit this fact in order to cancel the
residual anomaly. We are going to study two mechanisms in particular \cite{BCR}.
Other mechanisms have been considered in \cite{alwis},\cite{BCR}, but they fail
in one respect or another.

\subsection{A topological counterterm}

We examine the possibility to cancel the anomaly of the M--5--brane 
by means of the counterterm,
\a
S_{top} = \int_Y \omega_7 .\label{WZ}
\b
where eq.(\ref{Y}) is understood.
Formally,
\a
\delta S_{top} = -\int_Y d{\cal A}_6 = - \int_W {\cal A}_6.\0
\b

However we must exert some care in order for this to make sense.
We have just recalled that $\omega_7(N)$, as defined in (\ref{p2N}), is a basic 
form in $W$, and so it vanishes identically. Therefore
in order to have a non--trivial counterterm, we must extend $\omega_7(N)$
to $Y$ in a non--canonical (constructive) way, because the only canonical 
extension of 0 is 0; this in turn requires that we extend the connection and,
before that, the normal bundle.

In summary, we must make sure that:
\begin{itemize}
  
\item [1.] the normal bundle over $W$ extends to a bundle over $Y$;
\item [2.] the connection on the normal bundle extends too;
\item [3.] the counterterm (\ref{WZ}) is independent of the choice of  $Y$.
\end{itemize}

Conditions 1 and 2 are needed to ensure that the counterterm (\ref{WZ})
makes sense. Let us analyze them first.

To this end the normal bundles need a more precise notation: 
$N(X,Z)$ will denote the normal bundle of $X$ embedded in $Z$. For instance, 
according to this new notation, $N\equiv N(W,Q)$.  
First we have 
\a
\left. i^* T Q\equiv  T Q\right\vert_W=  T W\oplus N(W,Q), \label{norm1}
\b
where $i\colon W  \to Q$ is the embedding and 
\a
\left.  T Q\right\vert_W = \left.  T Y\right\vert_W \oplus N(Y,Q).\0
\b
As for $Y$, it is not closed, still we have
\a
\left.  T Y\right\vert_{W}= T W\oplus N(W,Y) .\0
\b
It follows that 
\a
N(W,Q)=\left. N(W,Y)\oplus N(Y,Q)\right\vert_W .\0
\b
Notice that $L \equiv N(W,Y)$ is a one-dimensional bundle. 
Since it is orientable, it is trivial and so we can write 
\a
N=N'\oplus L, \quad N'\equiv\left. N(Y,Q)\right\vert_W, 
\quad L= W\times {\bf R},\label{decomp}
\b
where $N'$ is an $SO(4)$-bundle, i.e. the gauge group $SO(5)$ reduces to 
$SO(4)$. The decomposition (\ref{decomp}) implies in particular that 
$p_2(N)=p_2(N')$.   
Since 
\a 
p_2(N')=\left. p_2(N(Y,Q))\right\vert_W\0, 
\b   
$p_2(N)$ extends to $0\in H^8(Y)$. As forms we can write
\a
p_2(N)=\left. e(N(Y,Q))\right\vert_{W}^2, \0
\b  
where $e$ denotes the Euler class.

After these preliminaries let us discuss the conditions 1 and 2 above.
Using (\ref{decomp}), one can trivially extend $L$ and $N'$ to bundles over 
$Y$ as follows:
\a
&&N'=\left. N(Y,Q)\right\vert_W \rightarrow N(Y,Q), \0\\ 
&&L =W\times {\bf R}  \rightarrow \tilde L = Y\times {\bf R}. \0
\b
Consequently $N$ extends in a natural way as
\a
N\rightarrow \tilde N = N(Y,Q) \oplus \tilde L. \label{ext}
\b
We can now extend the connection over this bundle as follows. Let us construct
a connection over $\tilde N$ by taking the trivial connection in $\tilde L$
plus the connection induced from $Q$ in $N(Y,Q)$. This is of course an
extension of the connection induced from $Q$ in $N$.
Therefore also the form $\omega_7$ extends, as well as all the operations on it.
All this is made possible by (\ref{decomp}).

It remains for us to discuss condition 3. By a standard argument the term
(\ref{WZ}) will not depend on the 
particular $Y$ 
if 
\a
{1\over {2\pi}} \int_X \omega_7 \in {\bf Z}\label{integer}
\b
for any compact 7--manifold $X$ without boundary.
Looking at (\ref{CS1}) it is actually easy to make a more precise statement
about the integral in (\ref{integer}). We remark that $\omega_7$ vanishes
when $A=A_0$. Now, a generic connection $A$ can be continuously joined to $A_0$.
Therefore also the results of the integral in (\ref{integer}) when evaluated
at $A_0$ and at $A$ should be continuously connected. This means that
the only value of the integral compatible with (\ref{integer}) is {\it zero}.

This condition limits the possible embeddings of the 5--brane.
It seems to be a rather strong one, but it is not easy to rewrite it in a
simpler form. Moreover, since the anomalies we are dealing with are 
field theory anomalies, it is natural to ask what is the relation of the
cancellation mechanism presented above with the locality principle. In fact
the term (\ref{WZ}) does not seem to be reducible to the locality principle,
\cite{BCR}. Therefore, if on one side we do not want to exclude a priori 
this mechanism, on the other side we can ask ourselves if the residual anomaly
can be cancelled by some kind of local counterterm. This is indeed possible,
as we will see in the next subsection.

\subsection{A world--volume counterterm}

We have seen at the beginning of the section that
a 5--brane in M--theory is characterized not only by the usual data (in
particular by eq.(\ref{mag})), but also by specifying a seven--manifold $Y$
that bounds its world--volume. However, if the 5--brane is to be specified 
by some other data beside the traditional ones, there is a less stringent
and more manageable way to do it than by specifying a bounding manifold 
$Y$: we can simply add the specification of a `collar' or `framing' 
attached to the 5--brane. 
The previous analysis told us that any 5--brane
is {\it framed}, i.e. that its normal bundle splits, as in (\ref{decomp})
-- with rather non--standard terminology we call such a splitting a framing.
If we consider a 5--brane with a {\it fixed} framing, we can construct a 
counterterm, akin to the one in \cite{witten1},
which, at some additional cost, cancels the anomaly.
We stress again that the fixed framing in this section is part of 
the definition of the 5--brane and is a local property as opposed to the
global existence of a bounding seven--manifold $Y$.

Therefore, let the normal bundle $N$ split as
\a
N=N' \oplus L, \label{C1}
\b
where $N'$ is an $SO(4)$-bundle and $L$ is  a trivial line bundle.
We have 
\a
p_2(N)=p_2(N') =e(N')^2 \label{pp1}
\b
and
\a
\Phi(N)=pr_1^*\Phi(N')\wedge pr_2^*\Phi(L),
\label{pp2}
\b
where $pr_i$, $i=1,2$ are the projections in the decomposition (\ref{C1})
of $N$ and $\Phi(E)$ is the Thom class \cite{bott} of a real 
vector bundle $E$ over $W$. $\Phi$ and $e$ are related by 
\a
\sigma_0^* \Phi(N') = e(N'),
\b 
where $\sigma_0\colon W\to N' $ is the zero section of the bundle $N'$. From
now on, for simplicity, we will drop the pull--back symbols.

Let $N_\epsilon(W)\subset N$ be a suitable tubular
neighborhood of $W$ of size $\epsilon$ embedded in $Q$.
If we identify $N_\epsilon(W)$ with the corresponding submanifold of $Q$,
then since  the support of $\Phi(N)$ can be taken to lie inside $N_\epsilon(W)$,
one can view $\Phi(N)$ as a form 
on $N_\epsilon(W)$ and identify it with the Poincar\'e dual  of
$W$, represented  by the form $\delta_W$.

Let $\omega_3$ be the basic 3--form constructed out of the polynomial
corresponding to $e(N')/24$ just like $\omega_7$ in eq.(\ref{CS1}).
Using the descent equations we get 
\a
\delta \omega_3 + d a=0, \0
\b
(\ref{pp1}) says that the integrated anomaly for the framed 
5--brane is given by 
\a
\int_W{\cal A}_6 = \int_W e' \wedge a.\0
\b
with $e' \equiv e(N')$. 

Let $v$ be  a nonvanishing vertical vector field on $L$ 
such that its contraction with $\Phi(L)$ satisfies
\a
\left. i_v \Phi(L)\right\vert_W =1. \label{v}
\b
The existence of such a field follows from the 5--brane being framed.
If $s$ is the coordinate along the fiber of $L$, we can simply choose
$v =\alpha \frac{\partial}{\partial s}$ with the constant $\alpha$ such that
(\ref{v}) be satisfied. 
If we now assume that the Lie derivative of $F_4$ in the $L$-direction
is zero on $W$,
\a
\left. {\cL}_v F_4\right\vert_W =0.\label{C2}
\b
then, using (\ref{mag}), (\ref{pp2}), (\ref{v}) and (\ref{C2}),
it is easy to see that
\a
\left. d i_v F_4\right\vert_W = - e' .\label{df4e}
\b
Now, consider the counterterm
\a
S= \int_W i_v  F_4\wedge \omega_3.\label{counter}
\b
Its gauge variation is, using (\ref{df4e}),
\a
\delta S= -\int_W  i_v F_4\wedge d a =
\int_W d i_v  F_4\wedge a = -\int_W e'\wedge a.\0
\b  
Hence the counterterm cancels the anomaly.

The case of IIA 5--brane, in which $Q=M\times S^1$, follows as a particular 
situation of the above. In this case the framing is determined by the 
background geometry.
 
Eq.(\ref{C2}), i.e. the constancy of $F_4$ along $L$ on $W$,
is the only condition required for this mechanism to work. It must be added to 
the defining eq.(\ref{mag}). We remark that the counterterm (\ref{counter})
depends on the vector field $v$, i.e. on the framing. In this section we have 
picked a definite one. However a 5--brane can have in general several distinct 
framings. In the previous subsection we saw that the 5--brane can have several
distinct $Y$'s that bound it. In this subsection we have seen that the global
datum of a bounding seven--manifold can be replaced by the local datum 
consisting of $v$, and the topological counterterm can be well replaced by a
local one. 

\section{Discussion}

In the above subsection
we have seen that the residual anomaly of the M--5--brane can be cancelled 
by the local six--dimensional counterterm (\ref{counter}) -- rather than
with the rather unappealing topological counterterm (\ref{WZ}).
However the anomaly cancellation problem does not end here. We have added a 
new degree of freedom, the vector field $v$. We have to ask ourselves 
whether this new degree of freedom is physical or not and whether it is 
dynamical or not. Unfortunately,
unlike in the Green--Schwarz mechanism for the superstring case, we do not know
at present the fundamental formulation of M--theory. Therefore we can only
formulate a few hypotheses, which are not inconsistent with what we know about
M--theory. If $v$ is
unphysical we should be able to soak it in some gauge symmetry. If it is 
physical, then the problem is how it should be understood. Perhaps the simplest 
interpretation is that $v$ represents a thickening of the 5--brane. 
In a related case, in \cite{aspin}, it was shown how a five--brane can be seen
as a limiting situation in F--theory, in which a squeezing takes place: in such
a case the world volume of the 5--brane can be regarded as higher 
dimensional manifold which becomes six--dimensional in the limit. If, as it is
reasonable, the 5--brane becomes anomalous only in the limit, the field $v$ 
could be interpreted as the reminder that the theory must be non anomalous.
In \cite{blo}, the authors introduced a (non--dynamical) vector field which
describes the background geometry of massive 11D supergravity and imposed 
conditions similar to (\ref{C2}). 

In any case a more complete formulation of $M$--theory with 
5--branes is necessary. Recently, on the wake of previous works, 
\cite{schwarz,pst}, a classical duality symmetric formulation of
11D supergravity with couplings to M-branes has been achieved in
\cite{BBS}. In such treatments bounding seven--manifolds for the 
five--brane world volume seem to be essential, while the gravitational 
correction term (\ref{bulk}) is not considered. It is clear that both issues 
are crucial for a comparison with our results.

Finally we should remark that the counterterm (\ref{counter}) breaks
supersymmetry. This reminds us again of the Green--Schwarz
mechanism in the LEEFT of the superstring; the Green--Schwarz counterterm
breaks supersymmetry. It is possible to recover supersymmetry by adding
an infinite number of terms to the equation of motions. However it is
impossible to satisfy all local field theory axioms: in particular, 
if we wish supersymmetry, locality and absence of anomalies in a field theory
with a finite number of fields, we are bound to find physical ghosts with
a consequent breaking of unitarity, see \cite{BBLPT} and references therein.
Of course this conclusion is not unespected since the theory in question is
a LEEFT, and represents superstring theory only approximately.
This fact is recalled here to suggest that even in the present case we may not
be so lucky as to find a closed solution of all the problems in a local
field theory action. After all M--theory with a 5--brane is much more complex
than superstring theories.

{\bf Acknowledgments} This paper is based on talks given by L.B. at the 
Conference on {\it Secondary Calculus and Cohomological Physics}, Moscow,
August 1997, at the 1997 European TMR Conference on {\it Integrability,
nonperturbative effects and symmetry in quantum field theory}, Santiago de
Compostela, September 1997, and at the Nordic Meeting on 
{\it Strings and M theory}, Uppsala, November 1997.
This research was partially 
supported by EC TMR Programme, grant FMRX-CT96-0012.

\end{document}